\documentclass[aps,twocolumn,epsf,floats,pre,nofootinbib]{revtex4-1}
\usepackage{graphics,graphicx,epsfig}
\usepackage{amssymb,color}
\usepackage{epsf,epstopdf,wrapfig}
\usepackage{amsmath}
\usepackage{amsmath}
\usepackage{amssymb}
\usepackage{graphicx}
\usepackage{bm}
\usepackage{xr}
\usepackage{scrextend}

\def\(({\left(}
\def\)){\right)}                       
\def\[[{\left[}
\def\]]{\right]}

\newcommand{\beq}{\begin{equation}}
\newcommand{\eeq}{\end{equation}}
\newcommand{\bea}{\begin{eqnarray}}
\newcommand{\eea}{\end{eqnarray}}

\newcommand {\lx} {\left}
\newcommand {\rx} {\right}

\newcommand {\Cw} {C(\omega)}
\newcommand {\Fw} {F(\omega)}

\newcommand {\PwB} {\prod_{i=1}^K (\omega - \omega_i)^{n_i} (\omega + \omega_i)^{n_i}}

\newcommand {\Res}[2] {\text{Res}\lx(#1,#2\rx)}

\begin{document}

\title{Dynamic scaling in natural swarms}

\author{
Andrea Cavagna$^{1}$, 
Daniele Conti$^{2}$,
Chiara Creato$^{1,2}$,
Lorenzo Del Castello$^{1,2}$, 
Irene  Giardina$^{1,2,3}$, \\
Tomas S. Grigera$^{4,5}$
Stefania Melillo$^{1,2}$, 
Leonardo Parisi$^{1,6}$,
Massimiliano Viale$^{1,2}$
}

\affiliation{$^1$ Istituto Sistemi Complessi, Consiglio Nazionale delle Ricerche, UOS Sapienza, 00185 Rome, Italy}
\affiliation{$^2$ Dipartimento di Fisica, Universit\`a\ Sapienza, 00185 Rome, Italy}
\affiliation{$^3$ INFN, Unit\`a di Roma 1, 00185 Rome, Italy}
\affiliation{$^4$ Instituto de F\'\i{}sica de L\'\i{}quidos y Sistemas Biol\'ogicos
%  (IFLYSIB), 
CONICET -  Universidad Nacional de La Plata, 
  %Calle 59 no.~789, B1900BTE 
  La Plata, Argentina}
\affiliation{$^5$ CCT CONICET La Plata, Consejo Nacional de Investigaciones Cient\'\i{}ficas y T\'ecnicas, Argentina}
\affiliation{$^6$ Dipartimento di Informatica, Universit\`a\   Sapienza, 00198 Rome, Italy}

\begin{abstract}
Collective behaviour in biological systems pitches us against theoretical challenges way beyond the borders of ordinary statistical physics. The lack of concepts like scaling and renormalization is particularly grievous, as it forces us to negotiate with scores of details whose relevance is often hard to assess. In an attempt to improve on this situation, we present here experimental evidence of the emergence of dynamic scaling laws in natural swarms. We find that spatio-temporal correlation functions in different swarms can be rescaled by using a single characteristic time, which grows with the correlation length with a dynamical critical exponent $z\approx 1$. We run simulations of 
a model of self-propelled particles in its swarming phase and find $z\approx 2$, suggesting that natural swarms belong to a novel dynamic universality class. This conclusion is strengthened by experimental evidence of non-exponential relaxation and paramagnetic spin-wave remnants,  indicating that previously overlooked inertial effects are needed to describe swarm dynamics. The absence of a purely relaxational regime suggests that natural swarms are subject to a near-critical censorship of hydrodynamics.
\end{abstract}

\maketitle

%\tableofcontents

%% Scaling laws in statphys
Scaling is one of the most powerful concepts in statistical physics. At the static level, the essential idea of the scaling hypothesis is that the only natural length scale of a system close to its critical point is the correlation length, $\xi$. In general, one could expect the behaviour of a system to depend in complicated ways on the parameters controlling its vicinity to the critical point.  The scaling hypothesis states that the situation is in fact simpler: the correlation functions depend on all these control parameters {\it only} through $\xi$ \cite{widom1965equation,kadanoff1966introduction}. The dynamic scaling hypothesis pushes this idea a step further by establishing a connection between space and time \cite{Ferrell1967, HH1967scaling}: when the correlation length is large, both the characteristic time scale and the dynamic correlation function depend on the control parameters only through the correlation length, which therefore becomes the sole relevant scale of the system also at the dynamical level. 
%Several relations connecting the various critical exponents descend from the scaling assumption, relations experimentally verified in a large variety of physical systems \cite{binney_book}. 
The dynamic scaling hypothesis is rooted in the renormalization group idea of studying how the laws of nature change under a rescaling of space and time. Close to criticality, scale invariance guarantees that all inessential microscopic details drop out of the quantitative description of a system. This is universality, the fundamental reason why a handful of physical laws have a vast range of applicability, from condensed matter to particle physics \cite{wilson1971renormalization1,wilson1971renormalization2}.

%% From physical to biological systems
The key ingredient of scaling is the existence of a large correlation length. This is not an exclusive prerogative of statistical physics.
Strong correlations are found in many biological systems composed by a large number of individuals; indeed, the very existence of significant correlations is arguably the best definition of collective behaviour \cite{attanasi2014collective}. Bird flocks \cite{cavagna+al_10}, fish schools \cite{strandburg2013}, mammals herds \cite{ginelli2015intermittent}, insect swarms \cite{attanasi2014collective}, bacterial clusters \cite{dombrowski2004self,zhang2010collective} and proteins \cite{tang2016critical} are all  biological systems where static correlations have been found to be strong. One may then wonder whether the concepts of scaling and universality make any sense in these contexts too.
A prudent answer would be negative: biological systems are characterized by a tumultuous balance between injection and dissipation of energy, and their complexity is far from our theoretical control. Yet one should  remember that even in statistical physics scaling is not a rigorous statement, but rather a phenomenological conjecture about what is relevant and what is not in a strongly correlated system. 
Hence, before ruling out scaling in the living world, one should test it experimentally.  Here we investigate the dynamic scaling hypothesis in natural swarms of insects. 
We find that dynamic scaling holds, and that a new and unexpected universality class emerges from the data.

\begin{figure*}
\centering
\includegraphics[width=1.00 \textwidth]{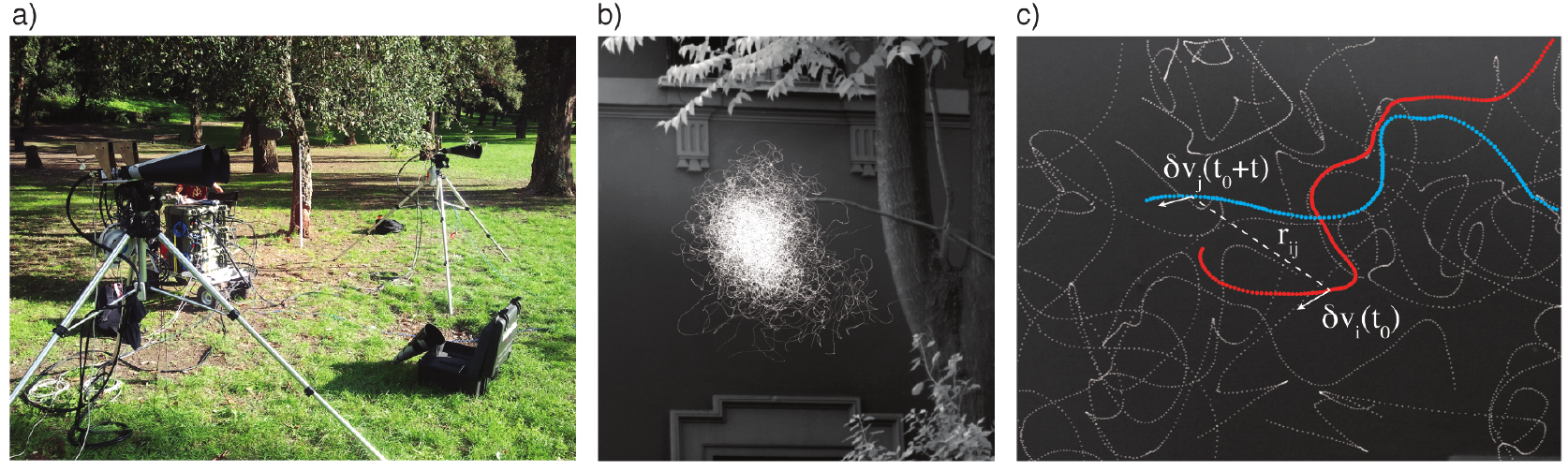}
\caption{ {\bf Experiment and correlation.} a) A system of three synchronized high-speed cameras (two on the left tripod, one on the right tripod) shooting at $170$fps is used to collect video sequences of midge swarms in their natural environment. b) A swarm of $N\sim 300$ midges: individual trajectories are reconstructed via the $3d$ tracking algorithm GReTA described in \cite{attanasi2015greta}. c) Close-up of two trajectories from the swarming event shown in panel b. The spatio-temporal correlation function $C(r,t)$ measures how much the velocity fluctuation of one insect at time $t_0$ influences the fluctuation of another insect a distance $r$ at time $t_0+t$.
}\label{fig::experiment}
\end{figure*}

%% Definition of the correlation function 
By using multi-camera techniques \cite{attanasi2015greta}, we reconstruct individual $3d$ trajectories in swarms of midges in their natural environment ({\it Diptera:Chironomidae} and {\it Diptera:Ceratopogonidae}; Fig.\ref{fig::experiment} and Methods). To perform a dynamic analysis, we conducted a new data-taking campaign based on the experimental setup of \cite{attanasi2014collective}, reaching a total of 30 natural swarms of various sizes and densities (Table \ref{table}). After the pioneering works of \cite{okubo1981use,gibson1985swarming, ikawa1994method}, new generation experiments on swarms have been performed both in the laboratory \cite{kelley2013emergent, puckett2014searching}, and in the wild \cite{butail20113d, butail2012reconstructing, attanasi2014finite, attanasi2014collective}. Natural swarms are characterized by strong static correlations and near-critical behaviour: the correlation length is large compared to the interparticle distance and the susceptibility far exceeds that of a noninteracting system  \cite{attanasi2014finite, attanasi2014collective}.  Hence, natural swarms are an ideal biological testbed for scaling concepts. From the trajectories we compute the spatio-temporal correlation function of the velocity fluctuations in Fourier space, 
\beq
C(k,t) = 
\left\langle \frac{1}{N} \sum_{i,j}^N   \frac{\sin[k\,r_{ij}(t_0,t)]}{k\,r_{ij}(t_0,t)}  \delta\hat{\bf v}_i(t_0) \cdot \delta\hat{\bf v}_j(t_0+t)  \right\rangle ,
\nonumber
\eeq 
where $\delta\hat{\bf v}_i$ is the dimensionless velocity fluctuation of insect $i$ and the brackets indicate an average over the earlier time $t_0$; 
the distance between insects $i$ and $j$ at different times is $r_{ij}(t_0,t) = |{\bf r}_i(t_0)-{\bf r}_j(t_0+t)|$, where positions are calculated in the centre of mass 
reference frame (see Methods). $C(r,t)$, the real space counterpart of $C(k,t)$, measures to what extent the velocity change of an insect at time $t_0$ influences that of another insect at distance $r$, at a later time $t_0+t$ (Fig.1c). For a frequency analysis of lab swarms dynamics see \cite{puckett2015time, ni2015intrinsic}.

In Fig.2a we report the normalized correlation, $\hat C(k,t)\equiv C(k,t)/C(k,t=0)$, as a function of time in various natural swarms. 
Each swarm is characterized by different size, density, and possibly other environmental parameters not under our direct control; all these factors can potentially affect the  
temporal decay rate of the experimental correlation.  To calculate the characteristic time scale, $\tau_k$, we follow the classic definition of \cite{HH1969scaling},
\begin{equation}
\int^{\infty}_{0} \frac{dt}{t} \, \sin(t/\tau_k)  \hat C(k,t)= \pi/4  .
\label{tau}
\end{equation}
%which is the time domain equivalent of the definition of characteristic frequency, $\omega_k\sim 1/\tau_k$, as the half-width of the correlation function in $\omega$ space. 
For a purely exponential correlation, $\tau_k$ coincides with the exponential decay time, while for more complex functional forms, $\tau_k$ is the most relevant time scale of the system. Relation \eqref{tau} gives an estimate of $\tau_k$ that is more robust than simply crossing $\hat C(k,t)$ with a constant and more reliable than a fit, as it does not require a priori knowledge of the functional form of $\hat C(k,t)$.

%%%%%%%%%%%%%%%%%%%%%%%%%%%%%%%%%%%%%%%%%%%%%
\begin{figure*}
\centering
\includegraphics[width=0.95\textwidth]{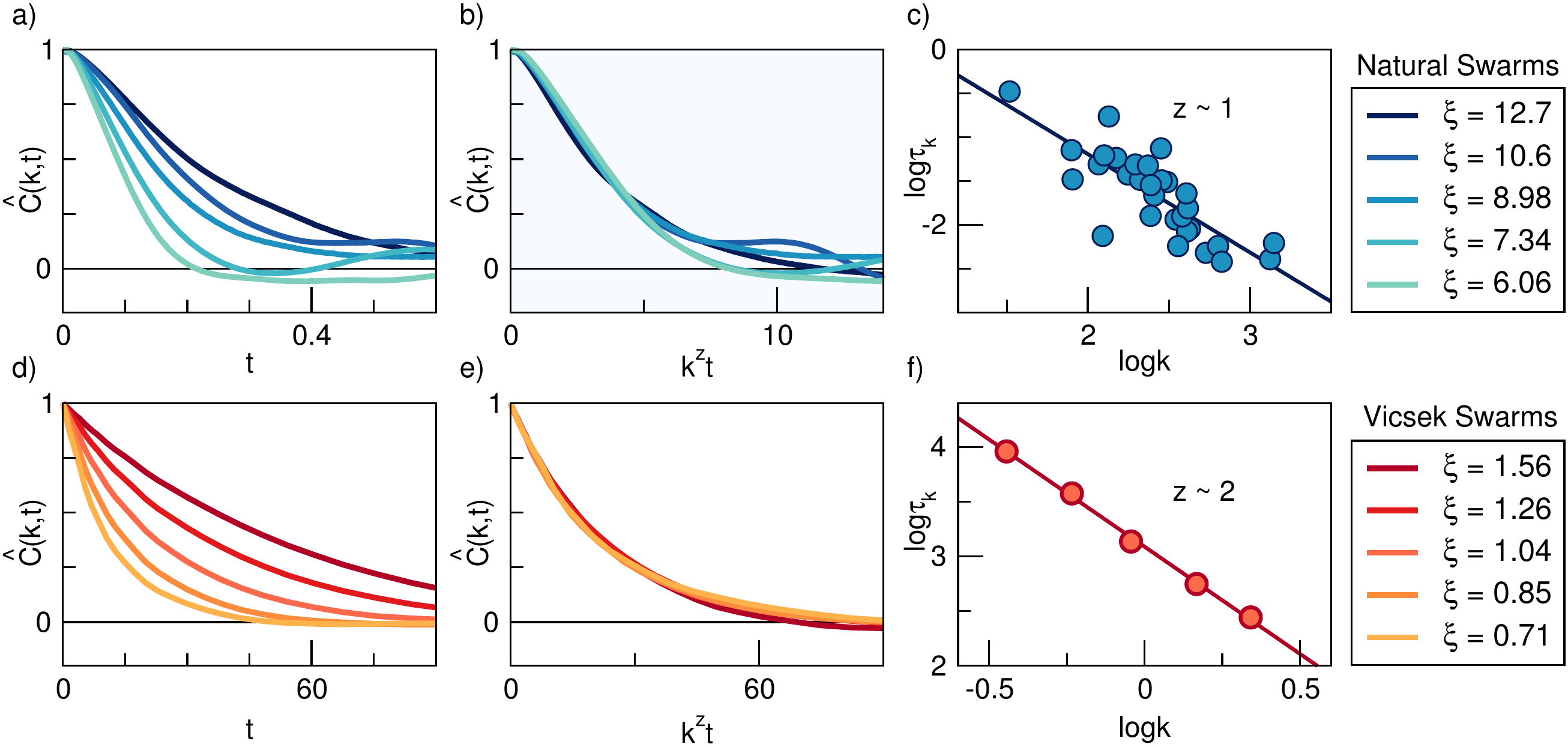}\\
\caption{{\bf Dynamic scaling and critical exponent.} 
a) Normalized time correlation function, ${\hat C}(k,t)$, evaluated at $k=1/\xi$, in several natural swarms. 
Sizes range from $N=100$ to $N=300$, time is measured in seconds and correlation length $\xi$ is centimeters.
b) ${\hat C}(k,t)$ as a function of the scaling variable $k^zt$ for the same events as in panel a); $z=1.2$ gives the optimal collapse of the curves according to equation \eqref{dsd2}. The quality of the collapse deteriorates for longer times because $\hat C(k,t)$ is the average over $t_\mathrm{max}-t$ time pairs ($t_\mathrm{max}$ is the sequence duration), hence large $t$ data are noisier. 
c) Characteristic time scale, $\tau_k$, computed at $k=1/\xi$, as a function of $k$ (log-log scale). Each point corresponds to a different natural swarm; all experimental events are reported. P-value=$10^{-6}$, $z=1.12\pm 0.16$, consistent with the estimate from the collapse in panel b). Panels d), e), f): dynamic scaling analysis of the $3d$ Vicsek model for $N=128, 256, 512, 1024, 2048$ particles; $\tau_k$ scale with $k$ with an exponent $z=1.96\pm 0.04$, which also produces an excellent collapse of the correlation functions.  In order to reproduce the phenomenology of natural swarms, density and noise have been chosen so that at each value of $N$ the system is at the maximum of the static susceptibility and the correlation length $\xi$ is proportional to the size $L$ (see \cite{attanasi2014finite} and Methods).  
}\label{fig::scaling_main}
\end{figure*}
%%%%%%%%%%%%%%%%%%%%%%%%%%%%%%%%%%%%%%%%%%%%%

% Dynamic scaling hypothesis
In absence of any general guiding principle, the time correlation function, $\hat C(k,t)$, and its characteristic time scale, $\tau_k$, could depend on the momentum $k$ and on the external parameters controlling the swarms' dynamics (density, noise, size, species, etc.) in complex ways. The dynamic scaling hypothesis  \cite{Ferrell1967, HH1967scaling, Ferrell1968, HH1969scaling} drastically reduces this complexity by conjecturing that both are simple homogeneous functions of $k$ and $1/\xi$,
\begin{align}
\hat C(k,t) &= f(t/\tau_k; k\xi)     , \label{ds2} \\
\tau_k &= k^{-z}  g(k\xi)   ,
\label{ds1}
\end{align}
where $f$ and $g$ are unknown scaling functions.
The fact that everything depends on the product $k\xi$ means that the correlation length, $\xi$, is the only quantity needed to locate swarms in their parameters space.  Eq.~\eqref{ds1} embodies the renormalization group idea that to a rescaling of space, $x\to x/b$, corresponds a rescaling of time, $t\to t/b^z$, a balance regulated by 
the so-called {\it dynamical critical exponent}, $z$ \cite{hohenberg1977theory}.

As first predicted in \cite{HH1967scaling}, a crucial consequence of the dynamic scaling hypothesis is that if we approach the critical point along paths of constant $k\xi$, a remarkable simplification in the structure of the time correlation emerges: correlation functions with different values of $k$ and $\xi$ must all collapse on the same curve, provided that time is scaled by $k^z$,
\beq
\hat C(k,t) = f(k^z t)   ,
\label{dsd2}
\eeq
while the characteristic time reduces to a simple power,
\beq
\tau_k \sim k^{-z} \sim \xi^z ,
\label{dsd1}
\eeq
where the last relation follows because $k\xi$ is kept fixed.  These two equations express the dynamic scaling structure that we are going to test in natural swarms.

% Experimental check of DS: timescale and collapse -  z=1
The simplest way to fix the product $k\xi$ in our data is to select $k=1/\xi$ in each swarm. There are several ways to evaluate the correlation length $\xi$ from the data (see Methods) and they all give the same results. 
The functional collapse predicted by Eq.~\eqref{dsd2} is reported in Fig.~2b. We find that the large spread of the correlation functions among different swarms is indeed significantly reduced when we rescale the time by $k^z$. The optimal collapse is obtained for $z=1.2$.
The characteristic time scale, $\tau_k$, at $k=1/\xi$, is reported in Fig.~2c. Although scatter is significant, the plot shows a clear correlation between $\log \tau_k$ and $\log k$ (P-value $\sim 10^{-6}$), in accordance with Eq.~\eqref{dsd1}; such correlation gives the dynamic critical exponent $z=1.12\pm 0.16$, consistent with the value of $z$ from the collapse.

% Piccolo pippotto moralistico su meaning of DS

Natural swarms therefore conform to a fundamental law of statistical physics: systems that are more {\it spatially} correlated (larger correlation length $\xi$), are also more {\it temporally} correlated (larger characteristic time $\tau_k$). This is the core of the dynamic scaling hypothesis: in a strongly correlated system, space and time are connected to each other by the exponent $z$. The fact that dynamic scaling holds in natural swarms is noteworthy for two reasons. First, these are off-lattice active systems, with a fiercely off-equilibrium nature; this suggests that scaling ideas have a reach that extends well beyond the borders of classic statistical physics. Secondly, the vicinity of swarms to their critical point is tuned by \emph{at least two} control parameters (noise level and density \cite{attanasi2014finite}, plus potentially many other biological and environmental factors we are unaware of), yet the correlation function is ruled by just {\it one} quantity, the correlation length. This fact strongly supports the idea that $\xi$ alone contains the most important effects of critical fluctuations \cite{HH1969scaling}.

%%%%%%%%%%%%%%%%%%%%%%%%%%%%%%%%%%%%%%%%%%%
\begin{figure*}
\centering
\includegraphics[width=0.80 \textwidth]{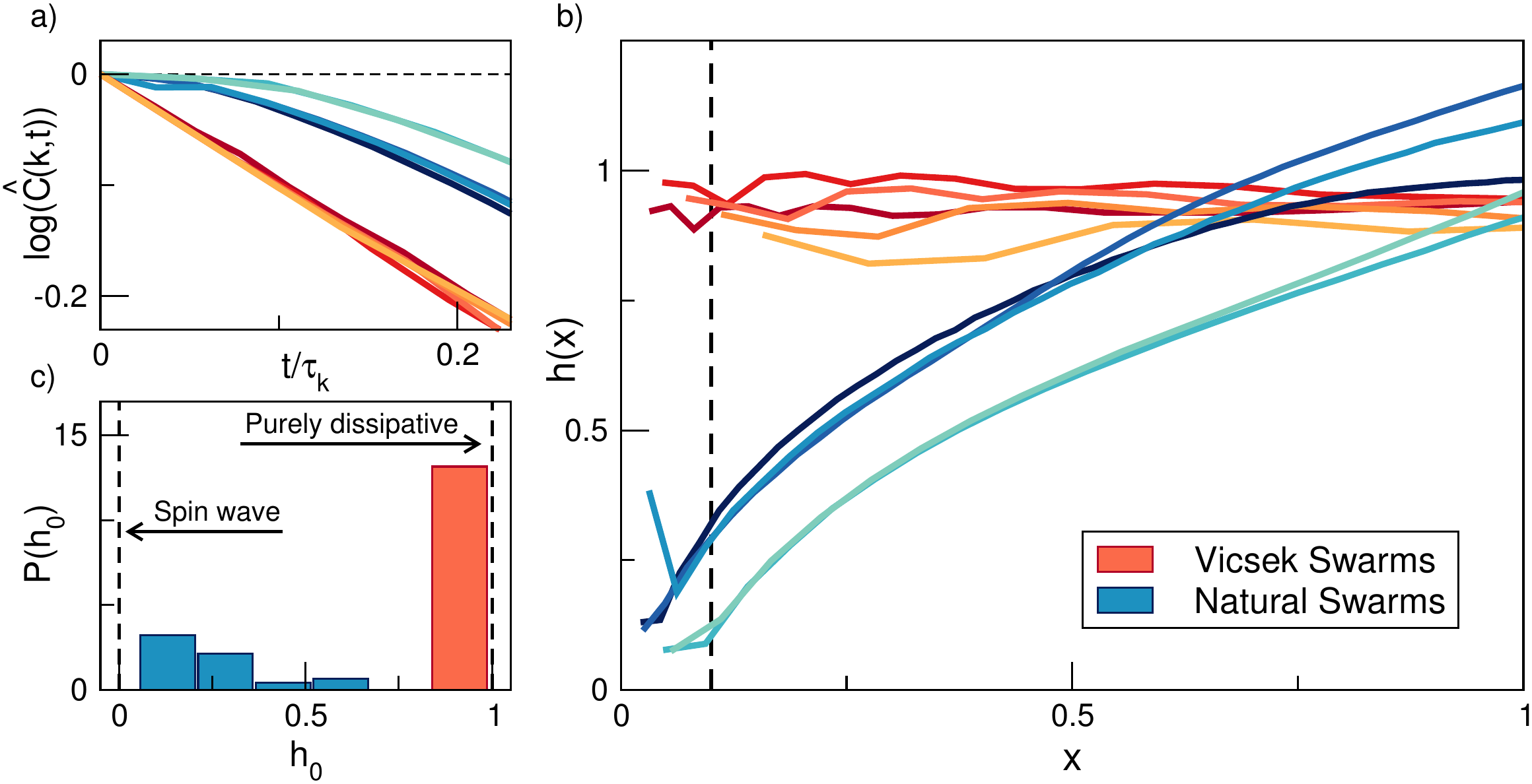}\\
\caption{{\bf Non-exponential relaxation and spin-wave remnant.}  a) Data of $\hat C(k,t)$ indicate that Vicsek swarms display exponential relaxation (linear decay in semi-log scale), while natural swarms have a strongly non-exponential correlation function (flat derivative for small $t$). b) To better quantify this difference we calculate the function $h(x)$ defined in \eqref{tosto}, where $x=t/\tau_k$; a clear difference emerges between natural and Vicsek swarms, the former being characterized by a small value of $h(x)$ in the whole interval $0<t< \tau_k$. c) In order to have a quantitative picture over all swarms, we compute the intercept $h_0=h(0.1)$ for all data and report its distribution: natural swarms have a strong non-exponential relaxation in the form of a low first derivative for small times, indicating the existence of spin-wave remnants. Vicsek swarms, on the other hand, have a clear peak at $h_0\sim1$, typical of purely dissipative dynamics.
}\label{fig::remnant}
\end{figure*}
%%%%%%%%%%%%%%%%%%%%%%%%%%%%%%%%%%%%%%%%%%%

% Vicsek: tau(\xi) and collapse - z=2
The value of $z$ determines the dynamical universality class of the system and it is therefore instructive to compare natural swarms ($z\approx 1$) to known theoretical models. The classical Heisenberg model of ferromagnetic alignment (Model A in the Halperin-Hohenberg classification \cite{hohenberg1977theory}) has $z\approx 2$; other non-dissipative magnetic models as Model G and Model J have $z=3/2$ and $z=5/2$, respectively \cite{hohenberg1977theory}. However, these are equilibrium lattice models far from the self-propelled nature of real swarms. A better term of comparison is  the Vicsek model of self-propelled particles \cite{vicsek+al_95}, which, in its near-critical phase, captures the static properties of natural swarms, in particular their density-dependent susceptibility \cite{attanasi2014finite}. The dynamic critical exponent of the Vicsek model near the ordering transition has been computed numerically in \cite{baglietto2008finite} in $d=2$, where it has been found $z\approx 1.3$. On the other hand, in the ordered phase the hydrodynamic theory of flocking \cite{toner+al_95, toner1998flocks}, consistently with numerical simulations \cite{kyriakopoulos2015leading}, predicts $z=2(d+1)/5$, namely $z=1.6$ in three dimensions. Hence, no estimate of $z$ exists in the literature for the Vicsek model in $d=3$ and in the swarm (disordered) phase. We run simulations of this case and find that the $3d$ Vicsek model in its near-critical paramagnetic phase satisfies dynamic scaling remarkably well (Fig.2d - Fig.2f - see Methods for details of the simulation). Both the collapse of the time correlations and the scaling of $\tau_k$ with $k$ give the dynamic critical exponent $z=1.96 \pm 0.04$, practically the same as classical Heisenberg, but twice as large as real swarms. We are not aware of other estimates of $z$ in different models or theories of swarm behaviour. The discrepancy between the dynamic critical exponent of natural swarms and that of all other $3d$ models, both on- and off-lattice, suggests that natural swarms belong to a potentially novel dynamic universality class. This opens intriguing new alleys for theoretical investigation.

%% Pippone Spin-Wave remnants

% Nonexponential relaxation: hard facts
A further hint that there is something qualitatively new in the dynamics of natural swarms comes from the shape of the time correlation function. While the Vicsek model displays plain exponential relaxation, real swarms have a clearly non-exponential correlation function, characterized by a vanishing first derivative for $t < \tau_k$ (Fig.3a). This feature seems at odds with the disordered nature of swarms and the seemingly dissipative motion of midges, both suggesting a purely diffusive dynamics of the velocity fluctuations, and thus exponential relaxation.  A concave correlation for $t<\tau_k$, on the other hand, is reminiscent of non-dissipative, inertial phenomena \cite{forster1975hydrodynamic}. It is therefore important to accurately verify this empirical result. To this aim we define the function,
\beq
%h(x)\equiv- \frac{\log \hat C(x)}{x} \quad , \quad x\equiv t/\tau_k \ ,
h(x)\equiv- \frac{1}{x} \log \hat C(x) \quad , \quad x\equiv t/\tau_k \ ,
\label{tosto}
\eeq
and study it in the interval $x\in[0,1]$, that is for times $t < \tau_k$. For purely exponential relaxation $h(x)\to 1$ for $x\to0$, while a flat time correlation gives $h(x)\to0$ in that same limit. We computed $h(x)$ in all swarms and find a very clear difference between natural and Vicsek swarms (Fig.3b,c), with the former showing a significantly lower value of $h(x)$ for $x<1$. We remark that this phenomenon emerges in the whole interval $t < \tau_k$, not for unnaturally short time scales.  The vanishing first derivative of the time correlation is thus a relevant trait of natural swarms over the time scales of interest, and not just a marginal feature.

% stacce
This type of non-exponential relaxation is a signature of the existence of propagating phenomena \cite{chaikin_lubensky, forster1975hydrodynamic}.
A vanishing first derivative of the time correlation function  can only arise if the dynamical propagator in the complex frequency plane has more than one pole (see Appendix A for proof), which in turns means that the dispersion polynomial is of degree two or more, so that the dynamical equation must involve second (or higher) time derivates.  Non-dissipative magnetic materials \cite{HH1969scaling}, superfluids \cite{hohenberg1977theory}, and bird flocks \cite{attanasi+al_14} are all systems characterized in their ordered phase by propagation of the fluctuations of the order parameter, namely by {\it spin waves} \cite{chaikin_lubensky}. In polarized animal groups, like flocks, propagating spin waves guarantee swift transmission of the velocity fluctuations, allowing the group to collectively change direction of motion while retaining cohesion \cite{attanasi+al_14}.  However, one would expect all propagating modes to be damped in the disordered (paramagnetic) phase, as is the case in swarms.  Actually, the fate of spin waves in the disordered phase depends on the product $k\xi$ \cite{HH1969scaling}: in the {\it hydrodynamic regime}, $k\xi \ll 1$, we are probing length scales much larger than the correlation length, so that spin-wave excitations are deeply damped and relaxation is exponential.  But if $k\xi \sim 1$, as in our data, we are in the {\it critical regime}: here we probe scales within a single correlated region, so that critical fluctuations invalidate the long-wavelength assumption of hydrodynamics. By far the most conspicuous hallmark of the failure of hydrodynamics in the critical regime is the possibility to have spin-wave excitations {\it even in the disordered paramagnetic phase} \cite{marshall1966critical, marshall1968magnetic, HH1969scaling}. The observable consequence in the time domain is a non-exponential temporal correlation with flat first derivative, which is what we find in natural swarms (Fig.3). We illustrate this phenomenon in a simple yet illuminating toy model in Appendix B (see in particular Fig.\ref{fig::scaling}).

% quasi-SW imply second order dynamics/inertia
The experimental evidence of spin-wave remnants in natural swarms strongly suggests that the equation underlying the collective behaviour of these systems admits propagating phenomena and therefore cannot be first order in time \cite{hohenberg1977theory}.  Swarms, like flocks \cite{attanasi+al_14,cavagna+al_15}, seem to be characterized by a second-order dynamics that is not captured by a purely dissipative first-order theory, which has purely exponential relaxation (Fig.~3).  It is tempting to speculate on how a new second-order swarm equation may look like. We see two main possibilities: i) propagating modes are chiefly caused by velocity fluctuations, implying the presence of non-dissipative terms and generalized inertia in the equation for the velocity, as in real flocks \cite{attanasi+al_14, hohenberg1977theory};  ii) propagating modes are the results of density waves coupled to velocity fluctuations, similar to the ordered phase of the Toner-Tu theory of active matter \cite{toner+al_95}. Even though the purely exponential relaxation we find in Vicsek swarms makes the second hypothesis less likely, it is hard to make a call purely based on our data. To make progress, it would be important to calculate the time correlation and the dynamic critical exponent $z$ in different \cite{couzin+al_02} and novel \cite{gorbonos2015long} theories of swarm collective behaviour and compare the results against the present experimental findings.

%%%%%%%%%%

% Near-critical censorship of hydrodynamics and the usefulness of dynamic scaling
Finally, let us remark that spin-wave remnants are found in the critical region, $k\xi\sim1$.  It would be natural then to expect hydrodynamics to take over and the correlation function to become exponential if we had examined the regime $k\xi\ll1$.  Interestingly, in natural swarms this is impossible.  Swarms are characterized by near-critical, scale-free spatial correlations, with a correlation length that scales with the system's size, $\xi \sim L$  \cite{attanasi2014finite}.  To access the hydrodynamic region we would therefore need $k\ll 1/L$, while the smallest accessible value is $k\sim 1/L$. We conclude that natural swarms are subject to a  near-critical censorship of hydrodynamics. Several biological systems are known to live in a near-critical regime \cite{mora+al_11} and may therefore share this same weird condition.
This scenario makes dynamic scaling particularly relevant for strongly correlated biological systems: by generalizing to non-equilibrium phenomena the usual scaling laws, dynamic scaling is not restricted to the hydrodynamic regime and can thus make predictions which fall outside the long-wavelength region, yet enjoy a high degree of universality even in finite-size near-critical systems \cite{HH1969scaling}. In particular, the dynamic critical exponent $z$ is {\it independent} of the specific regime (critical vs.\ hydrodynamic) and the dynamic universality class is therefore unequivocally identified. In natural swarms, $z\approx 1$ and spin-wave remnants are hard experimental benchmarks any future theory must confront with. Dynamic scaling may set equally useful benchmarks in other biological systems.

%%%%%%%%%%%%%%%%%%%%%%%%%%%%%%%%%%%%%%%%%%%%%%%%%%%%%%%%%%%%%%%%%%%%%%%
%%%%%%%%%%%%%%%%%%%%%%%%%%%%%%%%%%%%%%%%%%%%%%%%%%%%%%%%%%%%%%%%%%%%%%%
%%%%%%%%%%%%%%%%%%%%%%%%%%%%%%%%%%%%%%%%%%%%%%%%%%%%%%%%%%%%%%%%%%%%%%%
%%%%%%%%%%%%%%%%%%%%%%%%%%%%%%%%%%%%%%%%%%%%%%%%%%%%%%%%%%%%%%%%%%%%%%%
%%%%%%%%%%%%%%%%%%%%%%%%%%%%%%%%%%%%%%%%%%%%%%%%%%%%%%%%%%%%%%%%%%%%%%%

%%%%%%%%%%%%%%%%%%%%%%%%%%%%%%%%%%%%%%%%%%%%%%%%%%%%%%%%%%%%%%%%%%%%%%%
\section*{Methods}
%%%%%%%%%%%%%%%%%%%%%%%%%%%%%%%%%%%%%%%%%%%%%%%%%%%%%%%%%%%%%%%%%%%%%%%

%%%%%%%%%
{\bf Experiments.}
Data were collected in the field between May and October, in 2011, 2012 and 2015. We acquired video
sequences using a multi-camera system of three synchronized cameras (IDT-M5) shooting at 170 fps. 
%Two cameras (the stereometric pair) were at a distance between 3m and 6m
%depending on the swarm and on the environmental constraints. A third camera, placed at a distance of 25cm from the first camera was used to solve tracking ambiguities. 
We used Schneider Xenoplan 50mm f =2.0 lenses. Typical exposure parameters:
aperture f =5.6, exposure time 3ms. Recorded events have a time
duration between 1.5 and 15.8 seconds (see Table \ref{table}). More details can be found in \cite{attanasi2014collective}. 
To reconstruct the 3d positions and velocities of individual
midges we used the tracking method described in \cite{attanasi2015greta}. Our tracking method is accurate even on large moving groups and produces 
very low time fragmentation and very few identity switches, therefore allowing for accurate measurements of time-dependent correlations.
%(see Table \ref{table})
\vskip 0.5 truecm
%%%%%%%%%%
{\bf Correlation function.} 
We define the dimensionless velocity fluctuations as,
\beq
\delta \hat{\bf v}_i \equiv \frac{\delta{\bf v}_i}{\sqrt{\frac{1}{N} \sum_k \delta{\bf v}_k \cdot \delta{\bf v}_k}}  \ ,
\label{fluctu}
\eeq
where,
$\delta {\bf v}_i \equiv {\bf v}_i - {\bf V} $ and ${\bf V}$ is the collective velocity of the swarm which takes into account global translation, rotation and dilation modes, see \cite{attanasi2014finite}.
The spatio-temporal correlation function is the time generalization of the static space correlation function previously studied in \cite{cavagna+al_10, attanasi2014finite,attanasi2014collective},
%\cite{van1954correlations, cavagna2016spatio}, 
\beq
C(r,t) = \left\langle  
\frac{
\sum_{i,j}^N  \delta\hat{\bf v}_i(t_0) \cdot \delta\hat{\bf v}_j(t_0+t) \, \delta[r - r_{ij}(t_0,t)] 
}{
\sum_{i,j}^N  \, \delta[r - r_{ij}(t_0,t)] 
}
\right\rangle_{t_0} \ ,
\nonumber
\label{mingus}
\eeq
where $r_{ij}(t_0,t) = |{\bf r}_i(t_0)-{\bf r}_j(t_0+t)|$ and the positions are calculated with respect to the center of mass of the swarm, that is ${\bf r}_i(t_0) = {\bf R}_i(t_0) -{\bf R}_{\mathrm{CM}}(t_0)$; the brackets indicate an average over time,
\beq
\langle f(t_0, t)\rangle_{t_0}=\frac{1}{t_\mathrm{max}-t} \sum_{t_0=1}^{t_\mathrm{max}-t} f(t_0, t) \ ,
\eeq
where $t_\mathrm{max}$ is the total available time in the simulation or in the experiment. 
The purpose of $C(r,t)$ is to measure how much a  change of velocity of an individual at time $t_0$ influences a change of velocity of another individual at distance $r$ at a later time $t_0+t$. The (dimensionless) correlation function in Fourier space is given by,
\beq
C(k, t) = \rho \int d{\bf r} \ e^{i {\bf k}\cdot{\bf r}} C(r,t) \ .
\eeq
By using the definition of $C(r,t)$ and the approximation $\sum_{i,j}^N  \, \delta[r - r_{ij}(t_0,t)] \sim 4\pi r^2 \rho N$ in the integral, we obtain,
\bea
C(k,t) 
&&= \left\langle \frac{1}{N} \sum_{i,j}^N \;  \int_{-1}^{+1} d(\cos \theta) e^{ikr_{ij} \cos(\theta)}   \ \delta\hat{\bf v}_i \cdot \delta\hat{\bf v}_j \, \right\rangle_{t_0} 
\nonumber
\\
&&= \left\langle \frac{1}{N} \sum_{i,j}^N \;  \frac{\sin(k\,r_{ij}(t_0,t))}{k\,r_{ij}(t_0,t)} \ \delta\hat{\bf v}_i \cdot \delta\hat{\bf v}_j \, \right\rangle_{t_0} \ ,
\eea
which is the correlation function that we compute experimentally in the present work. Notice that, by definition, $\sum_i \delta \hat{\bf v}_i =0$; due to this sum rule we obtain $C(k=0,t)=0$. The smallest non-trivial value of the momentum we can evaluate the correlation at is therefore $k=2\pi/L$.

\begin{figure}
\centering
\includegraphics[width=0.4\textwidth]{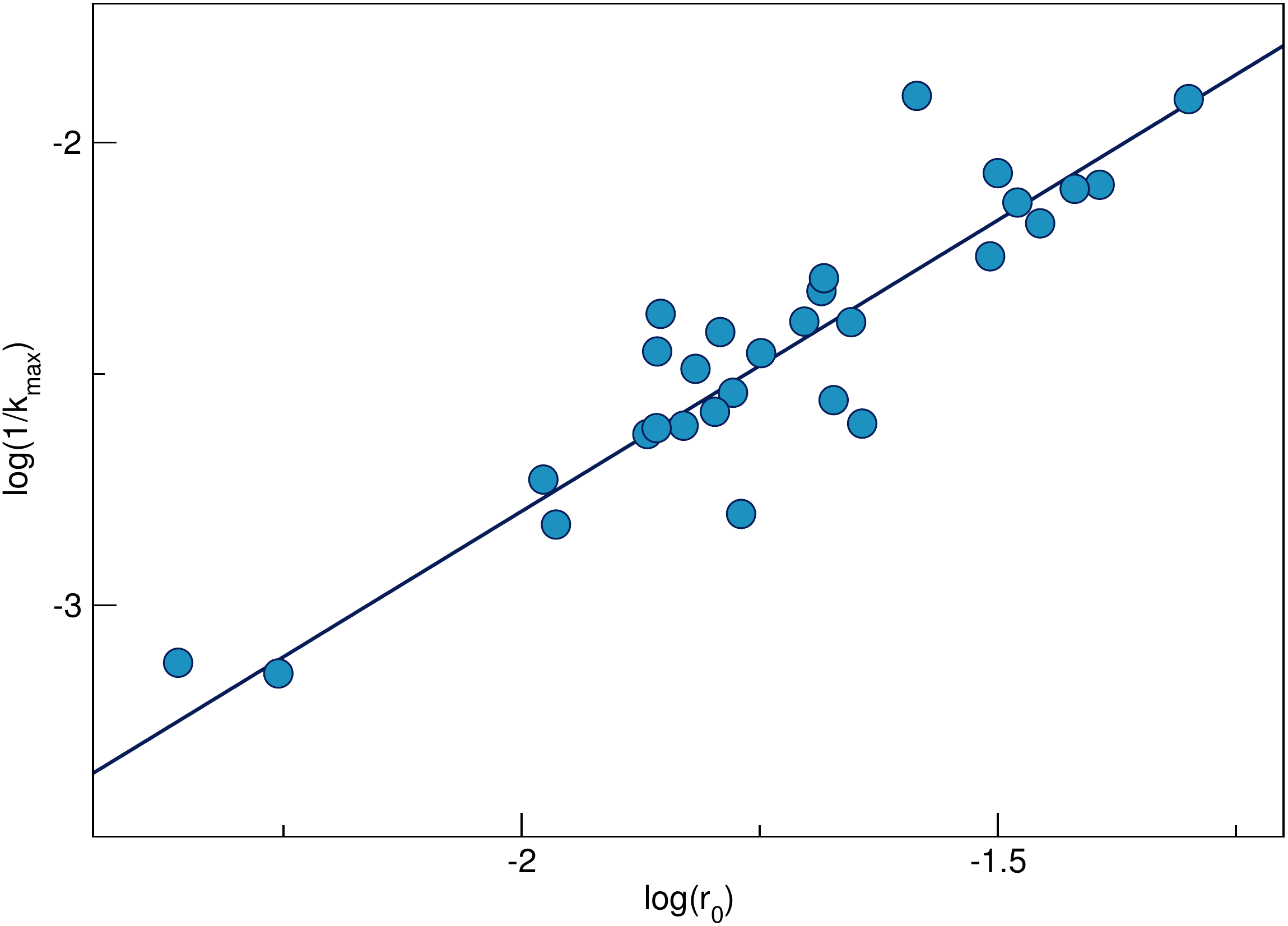}\\
\caption{
{\bf Correlation length.} The correlation length $\xi=1/k_{max}$ as a function of the correlation length $\xi=r_0$ computed from the static correlation function in r space as $C_0(r = r_0 = \xi)=0$ (log-log scale). Each point represents a different natural swarm. P-value $= 10^{-6}$, confirming the consistence of the two definitions of $\xi$.
}
\label{fig::xi}
\end{figure}

\vskip 0.5 truecm
%%%%%%%%%%%%%%
{\bf Correlation length.}
To compute the correlation length, $\xi$, we can directly work in $k$ space. The static correlation function, $C_0(k)\equiv C(k,t=0)$, is,
\beq
C_0(k)= \left\langle \frac{1}{N} \sum_{i,j}^N \;  \frac{\sin(k\,r_{ij})}{k\,r_{ij}} \ \delta\hat{\bf v}_i \cdot \delta\hat{\bf v}_j \, \right\rangle_{t_0} \ .
\label{galore}
\eeq
where now both $i$ and $j$ are evaluated at equal time, $t_0$. 
By decreasing $k$ we are averaging over larger length scales, therefore adding to \eqref{galore} more
correlated pairs, making $C_0(k)$ increase. 
When the momentum arrives at $k\sim 1/\xi$, we start adding uncorrelated pairs, hence, $C_0(k)$ must level. If we further decrease $k$ and reach $1/L$ (where $L$ is the system's size) we start to be affected by the sum rule, $C_0(k=0)=0$, hence the static correlation $C_0(k)$ decreases, until eventually it vanishes for $k=0$ \cite{cavagna2016spatio}.
In a system where $\xi \ll L$ the static correlation therefore has -- in log scale -- a broad plateau between $k\sim1/\xi$ and $k\sim1/L$. However, natural swarms are scale-free systems, where $\xi\sim L$ \cite{attanasi2014finite}; in this case, $C_0(k)$ has a well-defined maximum at $k_\mathrm{max} \sim 1/\xi \sim 1/L$. This is a very practical way to evaluate $\xi$ if one is already working in $k$ space and it is the one we use in this work.
Alternatively, one can define $\xi$ as the point where the static correlation in $r$ space, $C_0(r)=C(r,t=0)$ reaches zero, $C_0(r=\xi)=0$, as previously done in \cite{cavagna+al_10, attanasi2014finite,attanasi2014collective}. These two definitions of $\xi$ are consistent with each other (Fig.\ref{fig::xi}) and they both give the same dynamic scaling results.

\vskip 0.5 truecm
%%%%%%%%%%%%%%
{\bf Simulations.}  We simulated the Vicsek model \cite{vicsek+al_95} in $3d$ as
in \cite{attanasi2014finite}.  The update equations are,
\begin{align}
  \mathbf{v_i}(t+1) &= v_0 \mathcal{R}_\eta \left[ \sum_{j\in S_i} \mathbf{v}_j(t) \right],\\
  \mathbf{r_i}(t+1) &= \mathbf{r}_i(t) + \mathbf{v}_i(t+1),
\end{align}
where $S_i$ is a sphere of radius $r_c$ centred at $\mathbf{r}_i(t)$
and the operator $R_\eta$ normalises its argument and rotates it
randomly within a spherical cone centred at it and spanning a solid
angle $4\pi\eta$.  We chose $\eta=0.45$, $v_0=0.05$, $r_c=1$.

We considered systems of $N=128$, $256$, $512$, $1024$ and $2048$ particles, a range consistent with the typical sizes of natural swarms.  Dynamic scaling applies when $\xi$ is large, so we chose to have the largest possible $\xi$, i.e.\ to be at criticality.  This makes sense also because natural swarms are near-critical systems \cite{attanasi2014finite}.  To mimic the experimental situation, we fix the noise $\eta$ and use $x=r_1/r_c$ as control parameter, where  $r_1$ is the mean first-neighbour distance.  Scaling is then tested at pairs of values $(x,N)$ which lie along the critical line in the $x,N$ plane.  Note that $r_1$ cannot be fixed \emph{a priori,} but has to be determined from a simulation at a fixed average density.

For each value of $N$, several box sizes $L$ were chosen to obtain different average densities.  Five samples with random initial conditions were generated for each $N$ and $L$.  We ran each sample for $10^5$ steps for equilibration and used a further $5\cdot 10^5$ steps for data collection.  We verified that the polarisation $\Phi=(1/N)\sum_i \mathbf{v}_i/v_0$ remained stationary after the equilibration run, and that its correlation time was much shorter than $10^5$.  We then determined $r_1$ and computed the static correlation $C(k,t=0)$. This function has a maximum $C_\text{max}$ for some $k\equiv k_\text{max}$.  $C_\text{max}$ is a measure of the susceptibility $\chi$ (in statistical physics $\chi$ is given by the volume integral of $C(\mathbf{r},t=0)$, but in our case this integral is 0 because of the fact that $\sum_i \delta\mathbf{v}_i=0$ \cite{cavagna2016spatio}).  We thus obtained $\chi$ vs.\ $x$ curves from which we found the value of $x$ that maximises the susceptibility, $x_c(N)$: this is the finite-size critical point where the correlation length $\xi$ is of order $L$. We finally computed $C(k,t)$ at $x_c(N)$ (averaging over all samples) at $k=k_\text{max}(x_c(N))\sim 1/L$.  Since $\xi\sim L$, this fulfils the dynamic scaling condition $k\xi=\text{const}$ that we also adopt in natural swarms.

%%%%%%%%%%%%%%%%%
% References on scaling
%Patashinsky e Ma \cite{patashinskii_book, Ma_book}.
%To each static universality class are associated an infinite number of dynamic universality classes \cite{zinnjustin_QFTCF}.
%Dynamic scaling first formulated by Ferrel et al. \cite{Ferrell1967, Ferrell1968}, developed and rigorously formulated by Halperin and Hohenberg \cite{HH1967scaling, HH1969scaling, hohenberg1977theory}.
%%%%%%%%%%%%%%%%%

%%%%%%%%%%%%%%%%%%%%%%%%%%%%%%%%
\subsection*{Acknowledgements.}
We thank S. Caprara, F. Cecconi, F. Ginelli and J.G. Lorenzana for important discussions.  This work was supported by IIT-Seed Artswarm, European Research Council Starting Grant 257126, and US Air Force Office of Scientific Research Grant FA95501010250 (through the University of Maryland).  TSG was supported by grants from CONICET, ANPCyT and UNLP (Argentina).
%%%%%%%%%%%%%%%%%%%%%%%%%%%%%%%%

%%%%%%%%%%%%%%%%%%%%%%%%%%%%%%%%%%%%%%%%%%%%%%%%%%%%%%%%%%%%%%%%%%%%%%%%
%%%%%%%%%%%%%%%%%%%%%%%%%%%%%%%%%%%%%%%%%%%%%%%%%%%%%%%%%%%%%%%%%%%%%%%%
%%%%%%%%%%%%%%%%%%%%%%%%%%%%%%%%%%%%%%%%%%%%%%%%%%%%%%%%%%%%%%%%%%%%%%%%
%%%%%%%%%%%%%%%%%%%%%%%%%%%%%%%%%%%%%%%%%%%%%%%%%%%%%%%%%%%%%%%%%%%%%%%%

%%%%%%%%%%%%%%%%%%%%%%%%%%%%%%%%%%%%%%%%%%
%%%%%%%%%%%%%%%%%%%%%%%%%%%%%%%%%%%%%%%%%%
%%%%%%%%%%%%%%%%%%%%%%%%%%%%%%%%%%%%%%%%%%
\appendix

%%%%%%%%%%%%%%%%%%%%%%%%%%%%%%%%%%%%%%%%%%%%%%%%%%%%%%%%%%%%%%%

%%%%%%%%%%%%%%
\section{Structure of the correlation function in the complex $\omega$-plane.}

% poles
To interpret the non-exponential form of $\hat C(k,t)$ it is useful to reason in terms of the poles of its Fourier transform $\hat C(k,\omega)$ in the complex $\omega$-plane,
as their structure reflects the dispersion relation of the system and thus the underlying equation of motion 
\cite{lanczos1961linear}. What we will prove here is that exponential relaxation in time derives from a {\it single} pole of $\hat C(k,\omega)$ on the positive imaginary semi-plane, 
while a vanishing first derivative of the temporal correlation implies the existence of $\it two$, or more, poles of $\hat C(k,\omega)$ in the positive imaginary semi-plane.
From the Fourier relation,
\begin{equation} \label{e0}
  C(t) = \int_{-\infty}^{+\infty} d\omega \, e^{i \omega t} \Cw \ ,
\end{equation}
we have that the time derivative of the correlation function is given by,
\begin{equation} \label{eq1}
  \dot{C}(t) = \int_{-\infty}^{+\infty} d\omega \, e^{i \omega t} \Fw \quad , \quad \Fw = i \omega \Cw.
\end{equation}
From the physical condition $C(t)=C(-t)$, and therefore $C(\omega)=C(-\omega)$, we obtain that the
poles of $\Cw$ must have a symmetric structure,
\begin{equation}
  \Cw = \frac{1}{\PwB} \ , 
\end{equation}
where we admit that some pole may have multiplicity $n_i$ larger than one.

\begin{figure}[b]
  \includegraphics[width=0.4\textwidth]{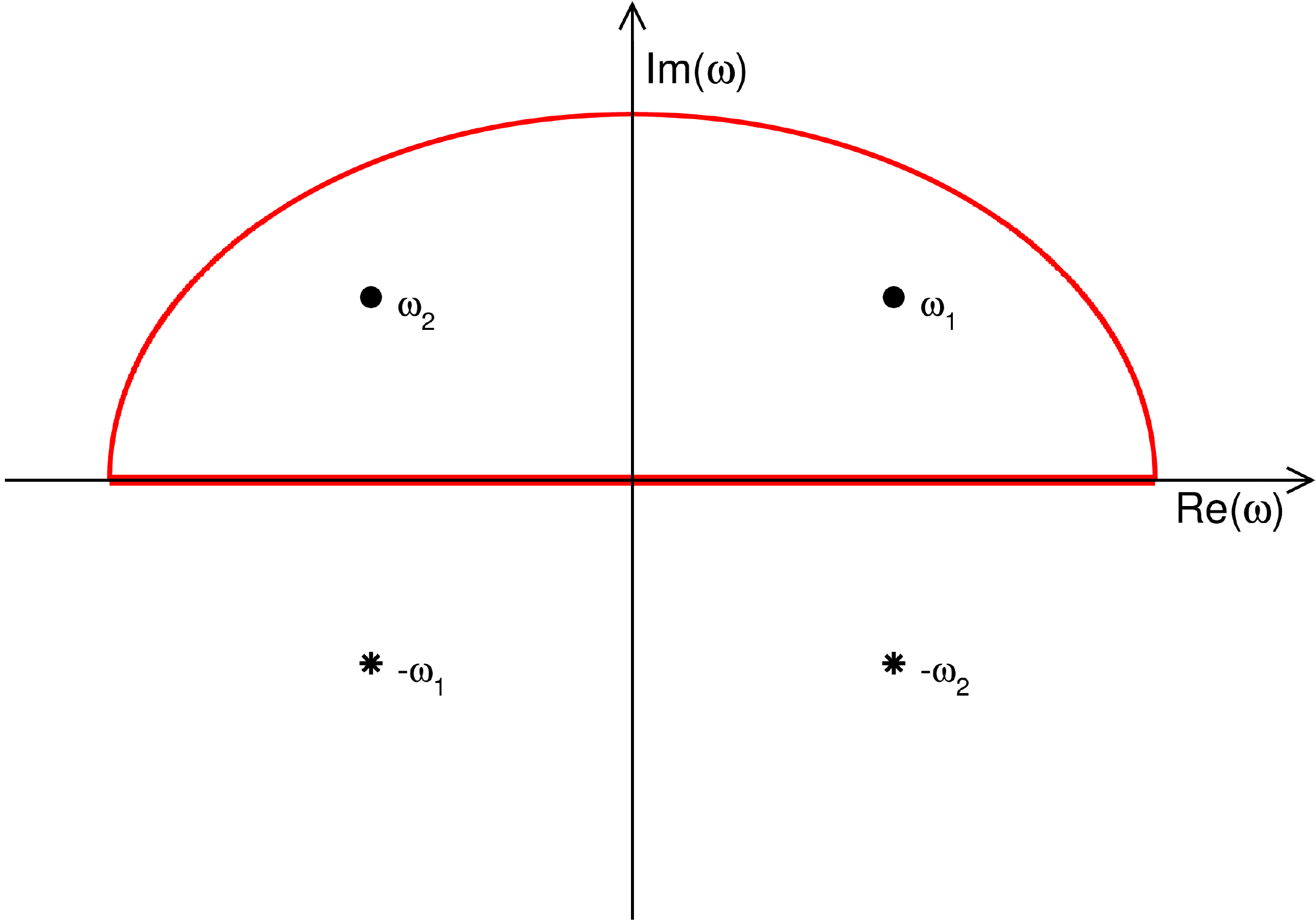}
  \caption{The integration path contains all poles of $\Cw$ with non-negative imaginary part (in this example we hypothesize that there
  are two such poles).}
  \label{fig:gaz}
\end{figure}

The $t \to 0^+$ limit of $\dot C(t)$ in \eqref{eq1} can be computed with the residue theorem by integrating $\Fw$ along the path 
in Fig.~\ref{fig:gaz}.
Because $F(-\omega) = -\Fw$, we have,
\begin{equation} 
  \Res{\Fw}{+\omega_i} = \Res{\Fw}{-\omega_i} \qquad \forall i=1,\dots, K \nonumber
\end{equation}
so that, after some algebra, we obtain,
\begin{equation} \label{eq2}
  \lim_{t \to 0^+}\dot{C}(t) = \frac{1}{2} \sum_{i=1}^K \lx[\Res{\Fw}{+\omega_i} + \Res{\Fw}{-\omega_i}\rx]
\end{equation}
The sum of all the residues of $\Fw$ coincides with its residue at infinity, $\Res{\Fw}{\infty}$, which can be computed as the residue in $z=0$ of the function $\hat{F}(z) = F(1/z)/z^2$,
\begin{equation}
  \Res{\hat{F}(z)}{0} = \lim_{\epsilon \to 0} \oint_{\cal{C}(\epsilon)} dz \frac{z^{\lx(2\sum_i n_i - 3\rx)}}{\prod_i (z^2 - 1/\omega_i^2)^{n_i} \prod_i \omega_i^{2n_i}} \ ,
   \nonumber
\end{equation}\\
where $\cal{C}(\epsilon)$ is a circle of radius $\epsilon$ centered in the origin.
The integral above is easily calculated, so \eqref{eq2} becomes,
\begin{equation}
  \lim_{t \to 0^+} \dot{C}(t) = \Res{\hat{F}(z)}{0} =
  \lx\{
  \begin{array}{ll}
    1 & \text{ if } \sum_i n_i = 1 \\
    0 & \text{ if } \sum_i n_i \geq 2 \\
  \end{array}
  \rx. \nonumber
\end{equation}
We conclude that a single pole in the positive semi-plane implies a non-zero first derivative of the time correlation function; more precisely, in 
this case $\Cw$ is a Lorentzian, so that $C(t)$ is purely exponential. On the other hand, a vanishing first derivative of the time correlation function $C(t)$ for $t\to 0$ (the feature we observe in natural swarms) is caused by the existence of two, or more, poles of its Fourier transform $\Cw$ in the positive imaginary semi-plane.

%%%%%%%%%%%%%%%%%%%%%%%%%%%%%%%
%%%%%%%%%%%%%%%%%%%%%%%%%%%%%%

The structure of these poles reflects the structure of the dispersion polynomial of 
the theory; in particular, multiple poles with a non-zero real part are the most distinctive hallmark of propagating spin-waves \cite{hohenberg1977theory}. In the overdamped, paramagnetic phase the real part of the spin-wave poles vanishes and the poles move onto the imaginary axis. Yet, their multiple structure (namely, the fact that they are more than one), remains as a remnant of the spin-wave phase and, as we have seen here, this remnant shows up as a zero derivative of the time correlation function. When we push
 a paramagnetic system deeply into its overdamped phase, i.e. down to the hydrodynamic phase, some of these poles becomes so large (high frequencies) that we no longer have the experimental resolution to see their effect in the derivative of $C(t)$, and we observe purely exponential relaxation. Natural swarms, though, are not in this phase, and a clear remnant of spin-wave poles is seen in the data.

%%%%%%%%%%%%%%%%%%%%%%%%%%%%%%%%%%%%%%%%%%
\section{Toy model of spin-wave evolution}
%%%%%%%%%%%%%%%%%%%%%%%%%%%%%%%%%%%%%%%%%%

Let us consider a classic system, the stochastic harmonic oscillator \cite{zwanzig_book},
\beq
m\,\ddot u(t) + \eta\, \dot u(t) + \kappa \, u(t)=\zeta(t)  \ ,
\eeq
where $u(t)$ is a generalized coordinate function of time, $m$ is the inertia, $\eta$ the viscosity and $\kappa$
the elastic constant, or stiffness. The noise has correlator,
\beq
\langle \zeta(t)\zeta(t')\rangle = 2 T\eta \delta(t-t') \ ,
\label{noise}
\eeq
where $T$ is the temperature.
To find the dynamic correlation of this linear stochastic equation it is convenient to consider the associate Green equation \cite{lanczos1961linear}, 
\beq
\left(m\frac{d^2}{dt^2} +\eta\frac{d}{dt} +\kappa \right) G(t-t') = \delta(t-t')
\label{prop}
\eeq
where $G(t-t')$ is the Green function, or dynamic propagator, of the theory. Once the dynamic propagator is known, the solution is given by (up to a solution of the homogeneous equation),
\beq
u(t) = \int dt' \ G(t-t')\zeta(t')
\eeq
and the time correlation function becomes,
\bea
C(t)&=& \langle u(t_0) u(t_0+t) \rangle =
\label{goss}
\\
&=& \int dt' dt'' G(t_0-t') G(t_0+t-t'')\langle \zeta(t')\zeta(t'')\rangle \ .
\nonumber
\eea
Using \eqref{noise} and passing in Fourier space of the frequency $\omega$ we get,
\bea
C(t)&=& 2T\eta\int dt' \  G(t_0-t')G(t_0+t-t') =
\nonumber
\\
&=& 2T\eta\int d\omega \ e^{i\omega t} G(\omega) G(-\omega) \ .
\label{zenga}
\eea
We therefore obtain the central relation between time correlation function and dynamic propagator in $\omega$ space,
\beq
C(\omega) = G(\omega) G(-\omega) \ .
\eeq
which is why $G(\omega$) is the central quantity in a stochastic theory. To calculate the dynamic propagator we rewrite
equation \eqref{prop} in Fourier space,
\beq
(-m \omega^2 + i\omega \eta +\kappa) G(\omega) = 1 , 
\eeq
which gives a simple algebraic expression of the dynamic propagator,
\beq
G(\omega) = \frac{1}{-m \omega^2 + i\omega \eta +\kappa} \ .
\label{propome}
\eeq
From \eqref{zenga} and \eqref{propome} we see that the form of the time correlation function $C(t)$ is entirely determined by 
the structure of the complex poles of the dynamic propagator $G(\omega)$, namely by the roots of the so-called {\it dispersion polynomial},
\beq
m \omega^2 - i\omega \eta -\kappa=0  \ .
\eeq
The stochastic differential equation we started from is of second order in time, hence the dispersion polynomial is quadratic. 
Once we introduce the two characteristic frequencies,
\beq
\omega_d = \eta/2m \quad , \quad \omega_0 =\sqrt{\kappa/m}  \ ,
\eeq
we can rewrite the dispersion polynomial as,
\beq
\omega^2 - 2i\omega\omega_d - \omega_0^2=0  \ ,
\eeq
which has the two roots,
\beq
\omega^{(\pm)} = i\omega_d \pm \sqrt{\omega_0^2-\omega_d^2}   \ .
\eeq
As we have seen, the dynamic propagator, $G(\omega)$, is the inverse of the dispersion polynomial and it therefore has two poles, $\omega^{(\pm)}$,
\beq
G(\omega) = \frac{1}{\omega-\omega^{(+)}} \cdot \frac{1}{\omega-\omega^{(-)}}  \ .
\eeq
The Fourier transform in \eqref{zenga} can be readily performed by using the residue method, to obtain the normalized time correlation function,
$\hat C(t)=C(t)/C(0)$,
\beq
\hat C(t) = e^{-\omega_d t}\left[  \frac{\omega_d}{\Delta\omega} \sin(\Delta\omega \; t) + \cos(\Delta\omega \; t)
\right]   \ ,
\label{explicit}
\eeq
where we have defined $ \Delta\omega = \sqrt{\omega_0^2-\omega_d^2}$.
The shape of the time correlation $\hat C(t)$ depends crucially on the damping ratio $\omega_d/\omega_0$. There are two regimes separated by a critical point
(see Fig.\ref{fig::scaling}):

%%%%%%%%%%%%%%%%%%%%%%%%%%%%%%%%%%%%%%%%%%%%%
\begin{figure*}
\centering
\includegraphics[width=0.75\textwidth]{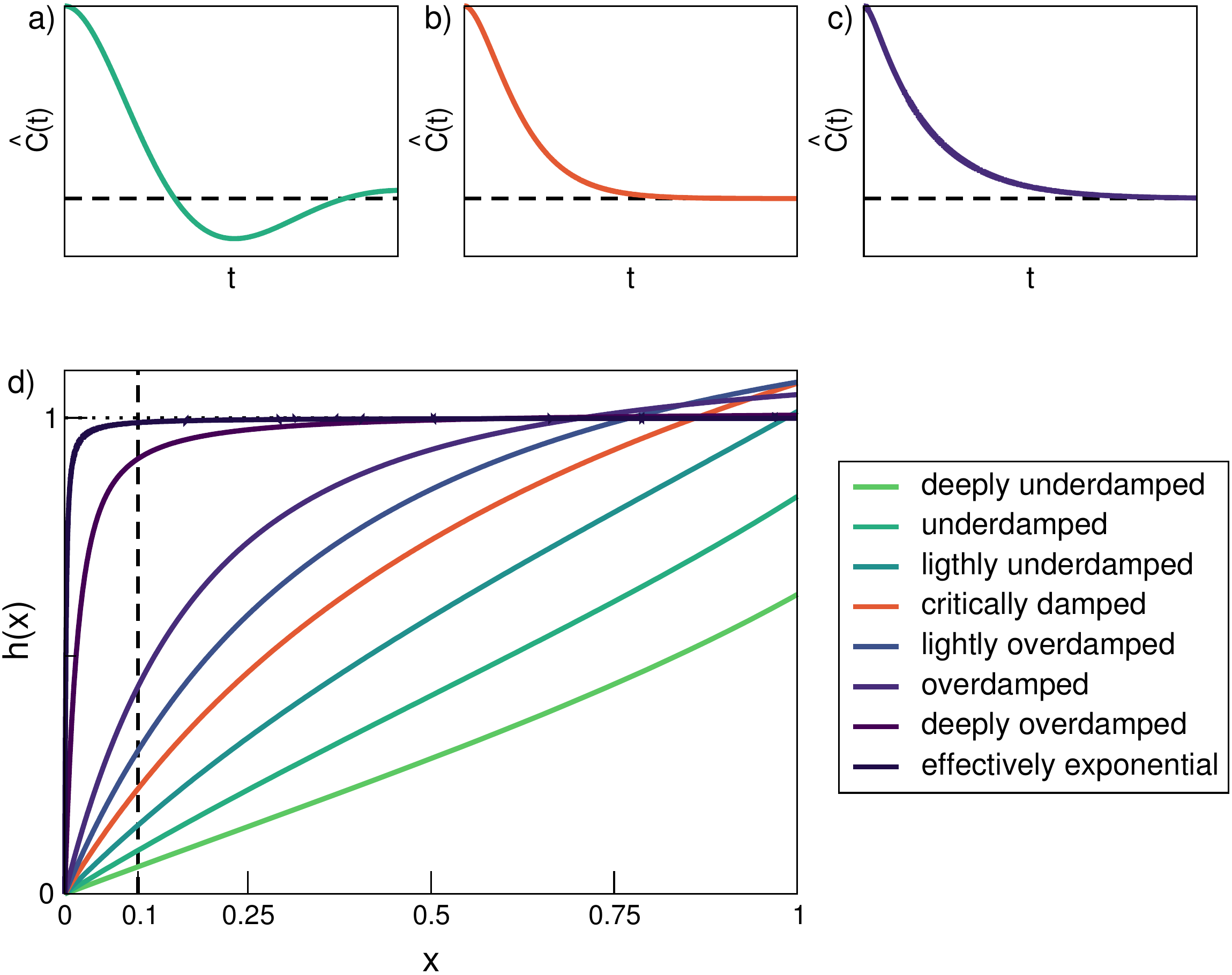}\\
\caption{{\bf Toy model of spin-wave remnant.}
Correlation function in the stochastic harmonic oscillator, eq.\eqref{explicit}, at different values of the damping ratio, $\omega_d/\omega_0$. a) Underdamped regime, $\omega_d/\omega_0 < 1$: the correlation function displays clear oscillatory behaviour and a flat first derivative for $t\to 0$. b) Critically damped, $\omega_d/\omega_0=1$: oscillations are no longer present, but a clear flat derivative for small times is still visible. c) Overdamped, $\omega_d/\omega_0>1$: the correlation function is more nearly exponential, even though non-exponential effects are still present for short times. d) The function $h(x)$ defined in \eqref{assuppa} clarifies how the correlation function crosses over from non-exponential ($h(x) \sim 0$ for $x\sim 0$) to pure exponential ($h(x)\sim 1$ for $x\sim 0$) as the damping grows. If the temporal resolution of our experiment is limited to $x>\epsilon$ ($\epsilon=0.1$ in the figure), when the damping ratio becomes very large the correlation function and the auxiliary function $h(x)$ cannot be distinguished from a pure exponential. On the other hand, the clear non-exponential form of the correlation is still clearly visible into the overdamped regime, as $h(\epsilon)\ll 1$; this non-exponential remnant in the overdamped phase corresponds to the spin-wave remnant in the paramagnetic phase of a system with alignment interaction.
}
\label{fig::scaling}
\end{figure*}
%%%%%%%%%%%%%%%%%%%%%%%%%%%%%%%%%%%%%%%%%%%%%

{\bf i)} For  $\omega_d/\omega_0 < 1$ we are in the
{\it underdamped} regime, where inertia (and stiffness) dominate over viscosity; the two poles have a large nonzero real part and a small
imaginary part, and the correlation function displays a clear oscillatory behaviour (Fig.\ref{fig::scaling}a); this regime corresponds to the propagating spin-wave phase
of ferromagnets \cite{hohenberg1977theory}.

{\bf ii)} At precisely $\omega_d/\omega_0 = 1$ the oscillator is {\it critically damped}, as inertia and viscosity exactly balance each other; the two poles have moved 
to the imaginary axis and coincide; the correlation function does not oscillate, and yet it retains a clear non-exponential form, with a flat correlation for small times (Fig.\ref{fig::scaling}b); 
the critically damped point  is the toy-model analogous of the critical point between ferromagnetic and paramagnetic phase. 

{\bf iii)}  For  $\omega_d/\omega_0 > 1$, the oscillator 
enters in the {\it overdamped} regime, where the time correlation becomes more and more exponential; the two roots are purely imaginary, 
$\omega^{(-)}\sim i\omega_0^2/2\omega_d\ll1$ and $\omega^{(+)}\sim 2i\omega_d$; this regime is the equivalent of the paramagnetic phase (Fig.\ref{fig::scaling}c).

Hence, by raising the damping of the oscillator, we have a modification of the time correlation function $\hat C(t)$ from an oscillatory, far-from-exponential behaviour in the underdamped regime, to a non-oscillatory, nearly-exponential behaviour in the overdamped regime, see Fig.\ref{fig::scaling}d. 

Yet it is straightforward to check from \eqref{explicit} that, irrespective of the regime we are in, the time correlation function {\it always} has vanishing first derivative for $t=0$,
\beq
\lim_{t\to 0^+} \frac{d\hat C(t)}{dt} = 0 \ .
\label{flat}
\eeq
This is a general result: when the dynamic propagator has more than {\it one} pole in the complex $\omega$-plane (and therefore $C(\omega)$ has more than one pole in the positive half-plane), the first derivative of $\hat C(t)$ in zero vanishes (we provide an explicit proof of this theorem in Appendix A); the stochastic equation we are studying is of the second order in time, hence the dispersion polynomial is quadratic and the propagator has two poles, and thus equation \eqref{flat} always holds. However, this result may seem confusing at the physical level: \eqref{flat} is a clear hallmark of non-exponential time correlation function, so it would seem that $\hat C(t)$ is non-exponential in {\it all} regimes; on the other hand, we just said, and showed in Fig.\ref{fig::scaling}, that the deeper we get into the overdamped phase, the more exponential the time correlation function becomes. The resolution of this paradox will bring us to a clearer understanding of the concept of spin-wave remnant.

What happens for increasing damping can be understood by following the evolution of the function,
\beq
h(x) \equiv -\frac{1}{x}\log \hat C(x)  \quad , \quad x\equiv t/\tau  \ ,
\label{assuppa}
\eeq
that we also study in the main text. Because $\hat C(0)=1$, a zero first derivative of the correlation for $t\to 0$ implies, 
\beq
\lim_{x\to 0} h(x) = 0  \ .
\label{acca0}
\eeq
On the other hand, a purely exponential time correlation implies,
\beq
\lim_{x\to 0} h(x) = 1  \ .
\label{acca1}
\eeq
Therefore, when we say that by increasing the damping the correlation becomes more and more exponential, we actually mean that the system crosses over from \eqref{acca0} to \eqref{acca1}. How this practically happens? The answer is clearly displayed in Fig.\ref{fig::scaling}: by increasing the damping, even though $h(x)$ is {\it always} zero at exactly $x=0$, the value of $x=t/\tau$ where $h(x)$ departs from $1$ becomes smaller and smaller. Our experimental apparatus must have a finite time resolution, as it is unphysical to think to be able to resolve the correlation for $x=t/\tau$ arbitrarily small; let us say that this experimental resolution is $t/\tau=\epsilon$, so we do not resolve time correlations for $t<\epsilon\, \tau$. This means that beyond a certain damping we are doomed to observe $h(x\sim \epsilon) \sim1$ within our experimental resolution, and the time correlation becomes therefore purely exponential for all practical purposes; this is what happens in the deeply overdamped phase (Fig.\ref{fig::scaling}). On the other hand, around the critically damped point and also in the weakly overdamped regime the departure from the exponential case is strong: the limit of $h(x)$ form small $x$ is clearly far from $1$ even within our experimental resolution $x>\epsilon$, and a clear flat derivative of the time correlation is experimentally visible. This is the mechanism underlying the existence of paramagnetic spin-wave remnant: although all the explicit oscillatory phenomena of spin waves are absent, the strong non-exponential character of the correlation function {\it in the experimentally relevant time regime $t\sim \tau$} is clear evidence that the original equation of motion admits spin-waves in a certain region of the parameters space and it is therefore second order in time.

%%%%%%%%%%%%%%%%%%%%%%%%%
%%%%%%%%%%%%%%%%%%%%%%%%%

\begin{table*}[b!]
\vskip 0.1 in
\begin{tabular}{|c|c|c|c|c|c|c}
\hline
\hline
 {\sc Event  label}   \hspace{0.2cm}     & \hspace{0.1cm}
 $N$    \hspace{0.2cm}     & \hspace{0.1cm}
 {\sc Duration (s)}    \hspace{0.2cm}     & \hspace{0.1cm}
 $\tau (s)$    \hspace{0.2cm}     & \hspace{0.1cm}
 $\xi (cm)$    \hspace{0.2cm}     & \hspace{0.1cm}
 $r_1 (cm)$    \hspace{0.2cm}     & \hspace{0.1cm}
 \\
\hline
\hline
20110511\_A2 & 279 & 0.88 & 0.12 & 12.3 & 5.33 \\
20110906\_A3 & 138 & 2.05 & 0.09 & 4.40 & 2.94 \\
20110908\_A1 & 119 & 4.41 & 0.11 & 4.30 & 3.59 \\
20110909\_A3 & 312 & 2.73 & 0.10 & 6.53 & 2.59 \\
20110930\_A1 & 173 & 5.88 & 0.47 & 11.9 & 5.72 \\
20110930\_A2 & 99  & 5.88 & 0.27 & 12.7 & 6.32 \\
20111011\_A1 & 131  & 5.88 & 0.23 & 14.9 & 7.52 \\
20120702\_A1 & 98 & 2.14 & 0.22 & 8.30 & 6.16 \\
20120702\_A2 & 111 & 7.29 & 0.14 & 7.88 & 5.57 \\
20120702\_A3 & 80 & 9.99 & 0.11 & 6.06 & 5.97 \\
20120703\_A2 & 167 & 4.41 & 0.09 & 5.93 & 4.65 \\
20120704\_A1 & 152 & 9.99 & 0.13 & 7.21 & 4.98 \\
20120704\_A2 & 154 & 5.29 & 0.13 & 7.34 & 5.32 \\
20120705\_A1 & 188 & 5.88 & 0.15 & 9.19 & 5.54 \\
20120828\_A1 & 89 & 6.29 & 0.11 & 7.75 & 6.18 \\
20120907\_A1 & 169 & 3.23 & 0.62 & 21.9 & 6.21 \\
20120910\_A1 & 219  & 1.76 & 0.24  & 10.6 & 4.68 \\
20120918\_A2 & 69 & 15.8 & 0.22 & 8.58 & 6.06 \\
20150729\_A1 & 110 & 5.87 & 0.32 & 8.61 & 4.63 \\
20150910\_A2 & 99 & 2.99 & 0.15 & 7.56 & 4.61 \\
20150921\_A1 & 201 & 4.11 & 0.23 & 9.81 & 4.21 \\
20150922\_A1 & 94 & 5.87 & 0.19 & 8.98 & 6.04 \\
20150922\_A2 & 126 & 5.87 & 0.29 & 11.4 & 5.29 \\
20150924\_A1 & 115 & 5.87 & 0.30 & 12.2 & 4.81 \\
20150924\_A4 & 107 & 4.38 & 0.32 & 15.0 & 6.38 \\
20151008\_A2 & 92 & 3.51 & 0.27 & 10.1 & 5.33 \\
20151008\_A3 & 91 & 5.87 & 0.16 & 7.30 & 4.41 \\
20151026\_A1 & 85 & 5.87 & 0.19 & 7.37 & 6.67 \\
20151030\_A1 & 274 & 5.87 & 0.27 & 9.34 & 3.96 \\
20151030\_A2 & 123 & 5.81 & 0.21 & 9.18 & 4.96 \\
\hline
\hline
\end{tabular}
\caption{Summary of experimental data. Swarming events are labelled according to experimental date and acquisition number. $N$ indicates the number of insects (and reconstructed trajectories) in the swarm. The correlation length $\xi$ is computed as $1/k_{\rm max}$, where $k_{\rm max}$ is the momentum where the static correlation has its maximum (see Methods). The characteristic time scale $\tau$ is computed following Eq. (1) of the main text, with $k=k_{\rm max}=1/\xi$. The behaviour of $\tau$ as a function of $k=1/\xi$ is displayed in Fig.~2c of the main text. The average nearest neighbour distance $r_1$ is calculated by averaging over all individuals in the swarm, and over the event duration. }
\label{table}
\end{table*}

%%%%%%%%%%%%%%%%%%%%%%%%%%%%%%%%%%%%%%%%%%%%%%%%%%%%%%%%%%%%%%%

\bibliographystyle{apsrev4-1}
\bibliography{general_cobbs_bibliography_file}

\end{document}